# Information-Theoretic Traffic Sensor Placement with Graph-Based Macroscopic Traffic-State Reconstruction

Ying Zhang[a], Fatemeh Fakhrmoosavi, Ph.D.[*b], Arash E. Zaghi, Ph.D.[c]

[a]*Graduate Assistant, School of Civil and Environmental Engineering, University of Connecticut, Storrs, CT, 06269, tjc25001@uconn.edu*

[b]*Assistant Professor, School of Civil and Environmental Engineering, University of Connecticut, Storrs, CT, 06269, moosavi@uconn.edu*

[c]*Professor, School of Civil and Environmental Engineering, University of Connecticut, Storrs, CT, 06269, arash.esmaili_zaghi@uconn.edu*

**Corresponding Author*

**ABSTRACT**
Traffic sensor placement in large urban networks must balance traffic monitoring needs with limited installation, maintenance, and data-management budgets. This study develops a framework for preserving network-wide traffic information while reconstructing flow and density at unobserved links and the macroscopic fundamental diagram (MFD). The framework uses normalized mutual information (NMI) to select sensors based on lag-aware joint flow–density dependence, information coverage, link importance, and redundancy, and an Inductive Graph Neural Network Kriging (IGNNK)-based graph-diffusion model to reconstruct traffic states at unobserved links. The framework was evaluated against a principal component analysis (PCA) benchmark using seven hours of dynamic traffic assignment simulation data for the Chicago network, with 4,805 directed links and 420 one-minute intervals split into training, validation, and independent testing subsets. PCA generally yielded lower root mean squared error (RMSE) over the 0.5%–2% range, but its errors were nonmonotonic and reached their minimum at 2%. At 3%, NMI-IGNNK outperformed PCA across all RMSE measures for both the testing subset and the complete analysis period, reducing normalized joint RMSE by 28.1% and 14.3%, respectively. NMI-IGNNK remained applicable up to 30% sensor coverage, with testing-subset normalized joint RMSE decreasing from 0.178 at 0.5% to its minimum of 0.046 at 10%. Increasing sensor coverage beyond 10% did not further reduce the normalized joint RMSE, likely reflecting diminishing marginal information once the major network traffic patterns were sufficiently represented. Sensitivity analyses at 3% and 10% sensor coverage showed that the effects of the sensor-selection hyperparameters varied with sensor budget.

## 1. Introduction

Traffic sensors are essential for monitoring and managing urban traffic conditions and supporting transportation planning and operations. However, limited budgets and constrained installation, maintenance, and data-management resources prevent permanent detectors from covering every urban link (Gentili and Mirchandani 2011; Owais 2022). In addition, the number of possible sensor-location combinations increases rapidly with network sizes, creating substantial computational complexity in large-scale transportation networks (Owais 2022; Yang et al. 2024). Therefore, effective algorithms are required to address the traffic sensor location problem in urban transportation networks.

Previous traffic sensor placement studies have addressed six major problem categories: origin–destination estimation and updating, flow observability, link-flow inference, path reconstruction, screen-line and traffic-surveillance design, and travel-time estimation (Owais 2022). Although sensors can be associated with either nodes or links, the majority of the studies focus on link-based placement. For link-flow applications, the Macroscopic Fundamental Diagram (MFD) provides a feasible means of aggregating complex link-level traffic conditions and estimating network-level performance (Saffari et al. 2020). The MFD represents the dynamic relationship between network-average flow and density, supports prediction of large-network behavior, and facilitates identification of network congestion states for traffic control applications (Godfrey 1969; Geroliminis and Daganzo 2008; Daganzo and Geroliminis 2008; Daganzo et al. 2011).

Existing MFD-based sensor-selection studies have generally optimized subsets of directly observed links to reproduce a complete or reference MFD. Although these approaches can identify representative links, their objectives are primarily based on aggregate MFD approximation and do not necessarily support reconstruction of flow and density at unobserved links. Principal component analysis (PCA)-based methods provide an alternative by identifying dominant network traffic patterns and reconstructing full-network states from critical-link observations (Saffari et al. 2020; El Bukhari et al. 2024). However, conventional PCA focuses on linear covariance structures and does not explicitly represent nonlinear, temporally lagged, or directionally asymmetric dependence. Because of the one-to-one correspondence between retained principal components and critical links, the number of independently supported sensor locations is constrained by the rank of the training matrix.

Mutual information (MI) provides an alternative measure of interlink dependence. MI quantifies the reduction in uncertainty about one random variable obtained by observing another and can characterize linear, nonlinear, monotonic, and nonmonotonic statistical associations (Shannon 1948; Cover and Thomas 2006; Kraskov et al. 2004). It can also support the selection of informative variables while limiting redundant observations (Krause et al. 2008; Peng et al. 2005). These properties are relevant to urban traffic networks, where signal control, queue formation and dissipation, time-delayed propagation, and directed roadway structures create complex relationships among traffic states on different links (Li et al. 2018).

This study develops an integrated normalized mutual information and Inductive Graph Neural Network Kriging framework (NMI-IGNNK). NMI is used to evaluate lag-aware dependence between joint flow–density states at candidate and target links. A greedy procedure selects sensors by balancing coverage of unobserved links, link-importance, and redundancy with previously selected sensors. The selected links then define a fixed observation mask for an IGNNK-based bidirectional graph-diffusion model that reconstructs flow and density at unobserved links and generates the corresponding MFD. The framework is evaluated using simulated traffic-demand data for the Chicago network and is compared with a PCA-based approach.

The contributions of this study are as follows: First, the study develops an NMI-based sensor-location method that represents lag-aware dependence between joint flow–density states and controls redundant selection among links with similar traffic patterns. Second, unlike PCA formulations that associate individual principal components with critical links, the NMI-based method is not strictly constrained by the number of retained principal components and can support a broader range of sensor budgets. Third, the selected sensor locations are incorporated into an IGNNK-based reconstruction model that represents network connectivity and traffic-movement relationships, producing both link-level traffic states and a network-level MFD. The method is evaluated on a large-scale Chicago network, demonstrating its applicability to complex sensor-location problems in real-world networks for which historical or simulation-based traffic data are available. Sensitivity analyses also reveal how the effects of sensor-selection hyperparameters vary across sensor budgets, providing guidance for budget-sensitive hyperparameter selection.

The remainder of this paper is organized as follows. The Literature Review discusses previous studies on transportation-network sensor selection, major methodological approaches, and applications of MI. The Methodology section presents the overall algorithm, network-data preparation, sensor-selection model, graph-aware reconstruction procedure, and evaluation metrics. The Case Study section describes the study network and traffic dataset. The Results and Discussion section presents and analyzes the MFD and root mean squared error (RMSE) results obtained from the NMI-IGNNK and PCA approaches under different sensor budgets. The Sensitivity Analysis section examines the effects of the four sensor-selection hyperparameters and discusses their implications for budget-sensitive parameter selection. Finally, the Conclusions section summarizes the findings, limitations, and directions for future research.

## 2. Literature Review

Previous studies have investigated how limited traffic measurements can be used to estimate the MFD (Geroliminis and Sun 2011; Zockaie et al. 2018). Ortigosa et al. (2014) examined the number and location of measurement points required for MFD estimation, while Zockaie et al. (2018) formulated a resource-allocation problem that jointly optimized fixed measurement locations and the sampling of probe trajectories under a limited data-collection budget. These approaches focused on estimating the aggregate MFD from partial network observations rather than reconstructing traffic states at unobserved links.

Beyond estimating the MFD directly from partial observations, data-driven dimensionality-reduction methods have been used to reconstruct network-wide traffic states from measurements at a limited number of locations. PCA is a low-dimensional representation technique that improves interpretability while minimizing information loss and has been widely applied to sensor placement and network reconstruction (Jolliffe and Cadima 2016). Saffari et al. (2020) applied PCA to identify dominant network traffic patterns and critical links, using observations from the selected links to reconstruct network-wide traffic states. Criticality coefficients were later introduced to aggregate information across principal components, while flow and density were considered jointly and spectral clustering was used to divide the network into homogeneous congestion regions (El Bukhari et al. 2024). PCA-based traffic-pattern identification focuses on linear variance and covariance structures and does not explicitly represent nonlinear or temporally lagged dependence. In addition, when each retained principal component is associated with a single critical link, the rank of the traffic-information matrix limits the number of independently supported principal components and sensor locations. This limitation can restrict higher-budget sensor selection when only a limited number of historical intervals are available.

These limitations motivate the use of information-theoretic measures that can characterize more general forms of dependence among traffic states. MI was introduced within information theory to quantify the reduction in uncertainty about one random variable obtained by observing another (Shannon 1948; Cover and Thomas 2006). Unlike linear correlation, MI can characterize general statistical dependence, provided that the joint distribution is estimated from sufficient data (Kraskov et al. 2004). MI-based selection can also balance relevance and redundancy by favoring informative observations while reducing duplicated information among selected variables (Krause et al. 2008; Peng et al. 2005).

The ability to capture nonlinear dependence is particularly relevant to urban transportation networks. Signal control, queue formation and dissipation, time-delayed propagation, and directed roadway topology can produce complex relationships among traffic states on different links (Li et al. 2018). Information-theoretic criteria have been applied to origin–destination sensor design and sparse traffic monitoring (Zhou and List 2010; Jusoh and Ampountolas 2019; Fei and Mahmassani 2011). Krause et al. (2008) established a theoretical foundation for greedy MI-based sensor placement using Gaussian processes, but their formulation characterized symmetric spatial dependence rather than directed traffic propagation. Fei and Mahmassani (2011) combined OD-demand coverage and uncertainty reduction using Kalman gains, eigenlinks, dynamic traffic assignment, and heuristic optimization, but the approach depended on prior OD and assignment information and focused on demand estimation rather than link-state reconstruction. Ivanchev et al. (2016) ranked intersections according to route-choice entropy and traffic volume, and robustness to OD variation was considered, while the method required modeled routes and turning probabilities and did not reconstruct continuous traffic states. Jusoh and Ampountolas (2019) used MI-based set covering and KL divergence to preserve the characteristics of MFDs under sparse measurements. Their approach considered only same-time flow–occupancy dependence and did not estimate traffic states at unsensed links.

Under limited sensor budgets, previous studies have used flow-propagation principles and statistical methods to extend observed sensor information to unmonitored portions of networks (Owais 2022). More broadly, data-driven and machine-learning methods have been applied to large-scale transportation networks to estimate unobserved or computationally expensive network quantities, including path travel costs (Xu et al. 2017; Rostami et al. 2024). In related large-scale transportation network design problems, clustering and heuristic optimization have been used to reduce the computational burden associated with large solution spaces (Kamjoo et al. 2024). Advanced machine-learning methods have been applied to traffic sensor location and reconstruction. Machine learning and sparse sampling have been combined for sparse sensor placement, signal identification, and reconstruction at unseen locations with available features; however, traffic reconstruction requires state estimation for known links whose dynamic features are unobserved (Manohar et al. 2018). Owais et al. (2019) developed a stacked sparse-autoencoder architecture to use traffic-flow data and network structure to extend information from a subset of links to the entire network. Gaussian-process-based and copula-based methods have also been combined with MI-based sensor selection (Krause et al. 2008; Zhang 2019). Wu et al. (2021) also developed IGNNK to recover signals at unsampled graph locations by learning spatial relationships from network topology and randomly masked traffic observations.

Building on this concept, the present study uses NMI-selected links as a fixed observation mask and applies bidirectional graph diffusion to reconstruct flow and density across unobserved links, where the NMI-IGNNK framework addresses these methodological gaps through lag-aware dependence between joint flow–density states, redundancy control, and directed graph-based reconstruction of traffic states at unobserved links.

## 3. Methodology

This study develops an integrated NMI-IGNNK framework for traffic sensor placement and network traffic-state reconstruction. The framework consists of three major components: joint flow-density NMI estimation, information-theoretic sensor selection, and graph-based reconstruction of network traffic states. First, joint flow-density NMI is estimated between candidate and target links to quantify spatial and temporal traffic-state dependence. Next, a greedy information-maximization algorithm iteratively selects sensor locations that maximize the weighted information coverage of the remaining unobserved links while reducing redundant information among selected sensors. Finally, an IGNNK-based graph diffusion neural network reconstructs traffic density and flow at unsensed links using observations from the selected sensors and the directed roadway topology. Summary of the NMI-IGNNK procedure is presented in Algorithm 1.

### 3.1. Network Representation and Joint Traffic-State Construction

The urban roadway network is represented as a directed link graph $G = (\mathcal{P}, \mathcal{E})$, where $\mathcal{P}$ is the set of directed roadway links and $\mathcal{E}$ denotes the set of feasible link-to-link movements. The traffic data consist of link density $k_i(t)$, expressed in vehicles per kilometer per lane (veh/km/lane), and total link flow $q_i(t)$, expressed in vehicles per hour (veh/h), for each link $i$ and time interval $t$. The time, link, and selected-sensor sets are defined as

$$\mathcal{T} = \{1, \dots, T\}, \qquad \mathcal{P} = \{1, \dots, P\}, \qquad \mathcal{S} \subseteq \mathcal{P}, \qquad |\mathcal{S}| \leq N \tag{1}$$

where $T$ is the number of time intervals, $P$ is the number of directed links, $\mathcal{S}$ is the set of selected sensor links and $N$ is the sensor budget. The link-level density and flow observations are organized into matrices $K \in \mathbb{R}_{\geq 0}^{T \times P}$ and $Q \in \mathbb{R}_{\geq 0}^{T \times P}$, respectively, where the $(t, i)$-th entries of $K$ and $Q$ are $k_i(t)$ and $q_i(t)$. To characterize the network-level traffic state at each time interval, the link-level observations are aggregated into network-average density and flow per lane. These quantities, defined below, represent the network-level traffic state and are used to construct the MFD (Fakhrmoosavi et al. 2020).

$$\bar{k}(t) = \frac{\sum_{i=1}^{P} k_i(t) l_i n_i}{\sum_{i=1}^{P} l_i n_i}, \qquad \bar{q}(t) = \frac{\sum_{i=1}^{P} q_i(t) l_i}{\sum_{i=1}^{P} l_i n_i} \tag{2}$$

where $l_i$ and $n_i$ denote the length and number of lanes of link $i$, respectively, and $l_i n_i$ is its lane length. Density is weighted by lane length because $k_i(t)$ is expressed per lane. Because $q_i(t)$ represents total link flow rather than per-lane flow, link length is used in the numerator of the network-flow expression, while total network lane length is retained as the denominator. The same aggregation definitions are applied to the observed and reconstructed link states to ensure direct comparability between their MFDs.

**Algorithm 1** NMI-IGNNK Sensor Placement and Fixed-Mask Traffic-State Reconstruction

**Input:** Directed link graph $G = (P, E)$; density matrix $K$; flow matrix $Q$; training, validation, and test interval sets $\mathcal{T}_{\text{tr}}$, $\mathcal{T}_{\text{val}}$, $\mathcal{T}_{\text{te}}$; sensor budget $N$; and model hyperparameters.
**Output:** Selected sensor set $S$ and reconstructed density and flow matrices $\hat{\text{K}}$ and $\hat{\text{Q}}$.

1: **procedure** NMI-IGNNK$(G, \text{K}, \text{Q}, \mathcal{T}_{\text{tr}}, \mathcal{T}_{\text{val}}, \mathcal{T}_{\text{te}}, N)$
2: Fit link-specific density and flow quantile thresholds using $\mathcal{T}_{\text{tr}}$
3: Construct joint flow-density states $z_i(t)$ for all $i \in \mathcal{P}$ and $t \in \mathcal{T}_{\text{tr}}$
4: Compute directed lag-aware coverage matrix C and symmetric same-time redundancy matrix R
5: Compute normalized target-link weights w
6: $\mathcal{S} \leftarrow \varnothing$; $c_j \leftarrow 0$ for all $j \in \mathcal{P}$
7: **for** $n = 1$ **to** $N$ **do**
8: Set $\mathcal{U} \leftarrow \mathcal{P} \setminus \mathcal{S}$
9: **for** each $i \in \mathcal{U}$ **do**
10: Compute the weighted marginal coverage gain $G_i(\mathcal{S})$ using Eq. (11)
Compute the link-importance using Eq. (12)
11: Compute the redundancy penalty using Eq. (14)
12: Compute the candidate score $\Delta_i(\mathcal{S})$ using Eq. (15)
13: **end for**
14: $i_{\text{best}} \leftarrow \text{argmax}_{i \in \mathcal{U}} \Delta_i(\mathcal{S})$
15: Add $i_{\text{best}}$ to $\mathcal{S}$
16: Update $c_j \leftarrow \max\{c_j, C_{j i_{\text{best}}}\}$ for all $j \in \mathcal{P} \setminus \mathcal{S}$
17: **end for**
18: Construct row-normalized forward and backward diffusion matrices $\text{P}_f$ and $\text{P}_b$ from the directed link topology, including self-loops
19: Estimate link–specific training means and standard deviations and construct the fixed sensor mask $m$
20: Construct split-specific temporal input tensors $\mathbf{X}_t$ with $2W + 1$ channels
21: Initialize the graph diffusion parameters θ and Adam optimizer
22: **for** epoch = 1 **to** $E_{\max}$ **do**
23: Shuffle training intervals and update θ over mini-batches by minimizing the total training loss in Eq. (19)
24: Evaluate the same loss function over $\mathcal{T}_{\text{val}}$
25: **if** the validation loss reaches a new minimum **then** save θ
26: **if** the validation loss does not improve within the specified patience period **then break**
27: **end for**
28: Restore the parameters with the minimum validation loss
29: Predict density and flow for all required intervals and transform the predictions back to physical units
30: Replace predictions at $i \in \mathcal{S}$ with the corresponding sensor observations
31: Verify that the reconstructed states are finite and nonnegative
32: Compute missing-link, all-link, and MFD reconstruction performance measures
33: **return** $\mathcal{S}, \hat{\text{K}}, \hat{\text{Q}}$
34: **end procedure**

For the information-theoretic analysis, the continuous density and flow observations are converted into discrete joint traffic states. Before discretization, the dataset is split into training, validation, and testing subsets, $\mathcal{T}_{\text{tr}}, \mathcal{T}_{\text{val}}, \mathcal{T}_{\text{te}}$, respectively. Link-specific quantile thresholds are estimated separately for density and flow using the training intervals. Let $B_K$ and $B_Q$ denote the maximum numbers of density and flow categories, respectively. For link $i$ at time $t$, let $b_i^K(t) \in \{0, \dots, B_K - 1\}$ and $b_i^Q(t) \in \{0, \dots, B_Q - 1\}$ denote the corresponding density- and flow-bin indices. The two indices are combined into a single joint flow–density state:

$$z_i(t) = B_Q b_i^K(t) + b_i^Q(t) \tag{3}$$

This encoding produces up to $B_K B_Q$ distinct states for each link. Multiplication of the density-bin index by $B_Q$ assigns a unique integer label to every possible density–flow bin combination. Therefore, two observations receive different joint-state labels whenever their density bins, flow bins, or both differ. The joint-state sequences from the training intervals are then used to estimate NMI between candidate sensor and target links.

### 3.2. Normalized Mutual Information

NMI is used to quantify the information shared between the joint flow–density states of candidate sensor links and target links (Cover and Thomas 2006; Strehl and Ghosh 2002). The information-theoretic formulation captures information sharing across the network while accounting for temporal delays and redundant information among candidate sensor locations, making it well suited for sensor placement in urban transportation networks (Zhou and List 2010; Denzler and Brown 2002). For candidate link $i$, target link $j$, and temporal lag $\tau$, the entropy of the joint traffic-state variable is defined as

$$H(Z_i) = -\sum_{a \in \mathrm{Z}} p_i(a) log p_i(a) \tag{4}$$

where $p_i(a)$ is the probability that link $i$ is in joint state $a$ and $Z$ denotes the set of all possible joint flow–density states (Strehl and Ghosh 2002). For temporal lag $\tau$, the mutual information between $z_i(t)$ and $z_j(t+\tau)$ is defined as follows.

$$I_{ij}^{(\tau)} = \sum_a \sum_b p_{ij}^{(\tau)}(a,b) \log\left[\frac{p_{ij}^{(\tau)}(a,b)}{p_i^{(\tau)}(a) p_j^{(\tau)}(b)}\right] \tag{5}$$

where $p_{ij}^{(\tau)}(a,b)$ is the joint probability that $z_i(t) = a$ and $z_j(t+\tau) = b$. The $p_i^{(\tau)}(a)$ and $p_j^{(\tau)}(b)$ are the corresponding marginal probabilities. The NMI is calculated as

$$\mathrm{NMI}_{ij}^{(\tau)} = \frac{I_{ij}^{(\tau)}}{\sqrt{H_i^{(\tau)} H_j^{(\tau)}}} \tag{6}$$

where $H_i^{(\tau)}$ and $H_j^{(\tau)}$ denote the entropies of the lag-aligned source and target state sequences, respectively. The NMI ranges from 0 to 1, where larger values indicate that the traffic state of candidate link $i$ provides more information about the traffic state of target link $j$. To quantify the maximum predictive information provided by each candidate sensor, a directed lag-aware coverage matrix is constructed as

$$C_{ji} = \max_{0 \le \tau \le \tau_{max}} NMI_{ij}^{(\tau)} \tag{7}$$

where the matrix $C \in \mathbb{R}^{P \times P}$ is directional with its rows as target links and its columns as candidate sensor links. Thus, $C_{ji}$ quantifies the information provided by sensor candidate $i$ for target link $j$. The diagonal is set to zero because a selected link is observed directly and is removed from the reconstruction-target set. To quantify duplicated information among candidate sensors, a same-time redundancy matrix is defined as

$$R_{ij} = max\left\{NMI_{ij}^{(0)}, NMI_{ji}^{(0)}\right\},\ R_{ii} = 0 \tag{8}$$

where the maximum of the two-directional same-time NMI values is used to construct a symmetric redundancy matrix, and the diagonal elements are set to zero to eliminate self-redundancy. The coverage and redundancy matrices provide the information-theoretic foundation for the sensor-selection procedure described later. Specifically, the coverage matrix is used to evaluate the weighted marginal information gain of each candidate sensor for the remaining unobserved links, whereas the redundancy matrix is used to penalize candidates whose information overlaps with that of previously selected sensors.

### 3.3. Sensor Selection

The sensor selection procedure aims to maximize the information available for reconstructing traffic states at unsensed links while minimizing redundant information among selected sensors. To accomplish this, a greedy information-maximization algorithm iteratively evaluates all unselected candidate sensors and selects the one that provides the largest increase in weighted information coverage after accounting for redundancy with the sensors already selected. The algorithm repeats this procedure until the specified sensor budget is reached. Because links contribute unequally to the network-level traffic state, each target link is assigned a normalized importance weight based on its lane length:

$$w_i = \frac{l_i n_i}{\frac{1}{P}\sum_{j=1}^{P} l_j\, n_j}, \qquad \frac{1}{P}\sum_{i=1}^{P} w_i = 1 \tag{9}$$

where $l_i$ and $n_i$ denote the length and number of lanes of link $i$, respectively. This normalization gives greater importance to links that contribute more to the network-wide traffic state while preserving an average weight of one. For a selected sensor set $\mathcal{S}$, the information coverage of an unobserved target link $j$ is defined as the largest lag-aware NMI between link $j$ and any selected sensor. In other words, each target link is considered to be represented by the selected sensor that provides the greatest amount of predictive information. The corresponding information coverage is:

$$c_j(\mathcal{S}) = \begin{cases} \max_{s\in\mathcal{S}} C_{js}, & \mathcal{S} \neq \emptyset \\ 0, & \mathcal{S} = \emptyset \end{cases} \tag{10}$$

When candidate sensor $i$ is evaluated, its weighted marginal coverage gain is

$$G_i(\mathcal{S}) = \sum_{j\notin\mathcal{S}\cup\{i\}} w_j\left[max\{c_j(\mathcal{S}), C_{ji}\} - c_j(\mathcal{S})\right] \tag{11}$$

which measures the additional information provided for reconstructing the remaining unobserved links. To measure the importance of the candidate link $i$, the link-importance contribution is defined as

$$O_i = \gamma\eta\pi_i \tag{12}$$

where the hyperparameter $\gamma$ is the link-importance coefficient, $\eta$ is a scale factor under sensor budget $N$, which was defined as $\eta = \frac{\sum_{j=1}^{P} w_j}{N}$, and $\pi_i$ is the composite link importance index, which was assigned to each candidate sensor, in account for normalized lane length, mean traffic activity, and temporal variability.

$$\pi_i = 0.50\,\tilde{L}_i + 0.25\,\tilde{A}_i + 0.25\,\tilde{V}_i \tag{13}$$

where $\tilde{L}_i$ represents normalized lane-length of candidate sensor $i$, $\tilde{A}_i$ represents normalized mean density and flow activity of candidate sensor $i$, and $\tilde{V}_i$ represents normalized density and flow variability of candidate sensor $i$ in the training subset.

To discourage the selection of sensors that provide overlapping information, the redundancy penalty is defined as

$$P_i(\mathcal{S}) = \alpha \sum_{s \in \mathcal{S}} R_{is} \tag{14}$$

where $\alpha$ is a hyperparameter that controls the influence of information redundancy. The candidate score, which serves as the greedy objective for selecting the next sensor, is calculated as

$$\Delta_i(\mathcal{S}) = G_i(\mathcal{S}) + O_i - P_i(\mathcal{S}) \tag{15}$$

Larger scores indicate that a candidate sensor provides greater additional information and has a higher link-importance across the network while contributing less redundant information. At each iteration, the candidate with the largest score is added to the sensor set, and the information coverage of the remaining unobserved links is updated accordingly.

$$i_{\text{best}} = \arg\max_{i \notin \mathcal{S}} \Delta_i(\mathcal{S}), \qquad c_j \leftarrow \max\{c_j, C_{j\,i_{\text{best}}}\}, \qquad j \in \mathcal{P} \setminus \mathcal{S} \tag{16}$$

The selected sensor is then removed from the reconstruction-target set because its traffic state is directly observed. This iterative procedure continues until the specified sensor budget $N$ is reached.

### 3.4. Graph Diffusion Traffic-State Reconstruction

After the sensor locations are determined, an IGNNK-based graph diffusion neural network is used to reconstruct traffic density and flow at the remaining unobserved links from the selected sensor measurements and the roadway network topology. The proposed model adapts the IGNNK framework to traffic sensor placement by incorporating a fixed sensor mask, a directed graph representation of the roadway network, bidirectional graph diffusion, and temporal traffic observations. The directed graph enables the model to capture spatial dependencies among connected roadway links, while the temporal input sequence incorporates traffic conditions from the current and preceding time intervals to improve reconstruction accuracy. The reconstructed link-level density and flow are subsequently aggregated to construct the corresponding MFDs.

*Graph Representation and Model Inputs*

A directed graph is constructed from the roadway network, where each node represents a roadway link and directed edges connect adjacent links according to feasible traffic movements. Row-normalized forward and backward diffusion matrices are then derived from the directed graph to model downstream and upstream information propagation, respectively. The bidirectional graph diffusion operation follows the original IGNNK formulation proposed by Wu et al. (2021). To improve numerical stability during model training, link-specific density and flow observations are standardized using statistics computed from the training subset:

$$\tilde{k}_i(t) = \frac{k_i(t) - \mu_i^K}{\sigma_i^K}, \ \tilde{q}_i(t) = \frac{q_i(t) - \mu_i^Q}{\sigma_i^Q} \tag{17}$$

where $\mu_i^K$ and $\mu_i^Q$ denote the training mean density and flow for link $i$, respectively, and $\sigma_i^K$ and $\sigma_i^Q$ are the corresponding standard deviations. When either standard deviation falls below a numerical tolerance, it is replaced by one to avoid numerical instability.

A fixed sensor mask is then applied so that traffic observations are available only at the selected sensor locations during both training and reconstruction. Let $m_i = 1$ if link $i$ is equipped with a selected sensor and $m_i = 0$ otherwise. Let $W$ denote the temporal input window length. For each time interval $t$, the graph neural network receives an input feature matrix $X_t$. The rows of $X_t$ correspond to roadway links, and its columns contain the masked standardized density and flow observations from the current and preceding $W - 1$ time intervals, together with the fixed sensor-mask vector. The input feature matrix is defined as

$$X_t = \left[m \odot \tilde{k}_t, m \odot \tilde{q}_t, \ldots, m \odot \tilde{k}_{t-W+1}, m \odot \tilde{q}_{t-W+1}, m\right] \in \mathbb{R}^{P\times(2W+1)} \tag{18}$$

where $\odot$ denotes elementwise multiplication, and $\tilde{k}_t$ and $\tilde{q}_t$ are the standardized density and flow vectors at time interval $t$. Consequently, density and flow observations are retained only at the selected sensor locations, whereas the remaining links are masked and reconstructed by the graph diffusion neural network. Following the IGNNK architecture, the model consists of two bidirectional graph diffusion layers with rectified linear unit (ReLU) activation, followed by a linear node-level output layer that estimates standardized density and flow for every roadway link (Wu et al. 2021). The predicted density and flow are subsequently transformed back to their physical units. Finally, the reconstructed values at the selected sensor locations are replaced with the corresponding observations, ensuring that measured traffic states are preserved while only unobserved links are reconstructed.

*Model Training and Optimization*

After the sensor locations are determined, the IGNNK-based reconstruction model is trained to estimate traffic density and flow at the remaining unobserved links. The model parameters are optimized using the Adam optimizer with randomly shuffled mini-batches from the training subset. Let $\mathcal{U} = \{i \in \mathcal{P}: i \notin \mathcal{S}\}$ denote the set of unobserved links. The model is trained by minimizing the total loss

$$L = L_{\mathcal{U}} + \beta L_{\mathcal{S}} + \lambda_{\text{MFD}} L_{\text{MFD}} + \lambda_{\text{phy}} L_{\text{phy}} \tag{19}$$

where $L_{\mathcal{U}}$ and $L_{\mathcal{S}}$ are the reconstruction losses for the unobserved and selected sensor links, respectively, $L_{\text{MFD}}$ is the network-level MFD reconstruction loss, and $L_{\text{phy}}$ is a physical penalty that discourages infeasible traffic-state predictions. The coefficients $\beta$, $\lambda_{\text{MFD}}$, and $\lambda_{\text{phy}}$ control the relative contributions of the corresponding loss terms. The reconstruction losses for both link subset $A \in \{\mathcal{U}, \mathcal{S}\}$ is computed as the lane-length-weighted mean squared error of the standardized density and flow predictions as follows.

$$L_A = \frac{\sum_t \sum_{i\in A} w_i \, \|\hat{y}_i(t) - y_i(t)\|_2^2}{2|\mathcal{T}_{\text{tr}}| \sum_{i\in A} w_i} \tag{20}$$

where $y_i(t)$ and $\hat{y}_i(t)$ denote the standardized ground-truth and predicted density-flow vectors, respectively. To improve reconstruction of the network-level traffic state, an additional MFD loss is also introduced in Eq. (21) as follows.

$$L_{\text{MFD}} = \frac{1}{|\mathcal{T}_{\text{tr}}|} \sum_{t\in\mathcal{T}_{\text{tr}}} \left[ \left(\frac{\Delta\bar{k}(t)}{s_{\bar{k}}}\right)^2 + \left(\frac{\Delta\bar{q}(t)}{s_{\bar{q}}}\right)^2 \right] \tag{21}$$

where $s_{\bar{k}}$ and $s_{\bar{q}}$ are the standard deviations of the network-average density and flow computed from the training subset, and $\Delta\bar{k}(t)$ and $\Delta\bar{q}(t)$ are the corresponding reconstruction errors calculated from the unobserved links using the same aggregation weights defined in Eq. (2). By optimizing both link-level and network-level errors simultaneously, the model improves reconstruction accuracy while preserving the macroscopic traffic-state characteristics.

To discourage physically infeasible predictions, a nonnegativity penalty is incorporated into the training objective:

$$L_{\text{phy}} = \frac{1}{|\mathcal{T}_{\text{tr}}| P} \sum_{t\in\mathcal{T}_{\text{tr}}} \sum_{i=1}^{P} \left[ \left(\frac{\text{ReLU}\left[-\hat{k}_i(t)\right]}{s_{\bar{k}}}\right)^2 + \left(\frac{\text{ReLU}\left[-\hat{q}_i(t)\right]}{s_{\bar{q}}}\right)^2 \right] \tag{22}$$

where $\text{ReLU}(x)$ is the maximum value between 0 and $x$. This penalty is zero for nonnegative predictions and increases quadratically when predicted density or flow becomes negative.

*Validation and Testing*

After each training epoch, the total loss is evaluated on the validation subset using the same objective function defined in Eq. (19). The model parameters corresponding to the lowest validation loss are retained throughout training, and optimization is terminated using an early stopping criterion when the validation loss fails to improve for a fixed number of consecutive epochs. The parameters associated with the minimum validation loss are then restored for the final model. Following training, the validation subset is also used to estimate a post-processing calibration that corrects systematic bias and range compression in the reconstructed MFD. The calibration is applied only to the reconstructed states of the unobserved links, and the resulting calibration coefficients are subsequently applied to all reconstructed traffic states. Finally, the reconstructed density and flow predictions are checked to ensure physically feasible traffic states. Specifically, the predicted density and flow are checked to ensure they both nonnegative.

$$\hat{k}_i(t) \geq 0, \qquad \hat{q}_i(t) \geq 0 \tag{23}$$

The final model performance is then evaluated using the independent testing subset.

### 3.5. Performance Evaluation Metrics

The reconstruction performance is evaluated by comparing the reconstructed and observed network-average traffic states using three RMSE measures: flow RMSE, density RMSE, and a joint normalized RMSE that combines both variables (Saffari et al. 2020). These metrics are used to compare the proposed NMI-IGNNK framework with the PCA-based sensor-placement method.

$$RMSE(Q) = \sqrt{\frac{1}{T}\sum_{t=1}^{T}(\hat{\bar{q}}(\mathrm{t}) - \bar{q}(\mathrm{t}))^2} \tag{24}$$

$$RMSE(K) = \sqrt{\frac{1}{T}\sum_{t=1}^{T}\left(\hat{\bar{k}}(\mathrm{t}) - \bar{k}(\mathrm{t})\right)^2} \tag{25}$$

$$RMSE(Q,K) = \sqrt{\frac{1}{T}\sum_{t=1}^{T}\left[\left(\frac{\hat{\bar{q}}(\mathrm{t}) - \bar{q}(\mathrm{t})}{Q_c}\right)^2 + \left(\frac{\hat{\bar{k}}(\mathrm{t}) - \bar{k}(\mathrm{t})}{K_c}\right)^2\right]} \tag{26}$$

where $\bar{q}(t)$ and $\bar{k}(t)$ denote the observed network-average flow and density at time interval $t$, respectively, and $\hat{\bar{q}}(t)$ and $\hat{\bar{k}}(t)$ are the corresponding reconstructed values. The normalization constants $Q_c$ and $K_c$ denote the maximum observed network-average flow and a characteristic upper bound for the network-average density, respectively. The latter is approximated by the mean of the three largest observed network-average density values to reduce sensitivity to extreme observations. The joint RMSE normalizes the flow and density errors so that both variables contribute comparably despite their different units and numerical ranges. All RMSE values are computed over the evaluation subset under consideration (training, validation, testing, or the complete dataset).

## 4. Case Study

### 4.1. Chicago Network and Simulation Data

The proposed framework was evaluated using the Chicago network, a large-scale urban roadway network consisting of 1,578 nodes and 4,805 directed links, as illustrated in Figure 1. The network was simulated in DYNASMART-P over a seven-hour period. Time-varying travel demand was loaded during the five-hour morning period from 5:00 a.m. to 10:00 a.m., followed by two additional hours without new demand loading to allow vehicles remaining in the network to complete their trips and congestion to dissipate (Fakhrmoosavi et al. 2023).

The simulation generated complete link-level traffic-state observations at one-minute resolution. For each directed link and time interval, the extracted data included traffic density, expressed in vehicles per kilometer per lane, and total link flow, expressed in vehicles per hour. The resulting dataset contained 420 one-minute intervals for each of the 4,805 links. Link length, number of lanes, and directed link-to-link connectivity were also retained to construct the roadway graph, calculate lane-length weights, aggregate link-level traffic states, and apply the physical plausibility checks described in the methodology.

The use of simulation data provided complete observations for both selected and unselected links throughout the analysis period. This full network coverage was necessary to estimate information dependence between candidate and target links, train the reconstruction model under controlled sensor masks, and compare reconstructed traffic states with known ground-truth values. In an operational setting, measurements would be available only at instrumented links. However, in this case study, the complete simulation output served as the reference against which alternative sensor configurations and reconstruction results were evaluated.

At each time interval, the network-average density and flow were calculated using the lane-length-weighted aggregation defined in Eq. (2). The resulting sequence of network-average traffic states was used to construct the observed MFD and to evaluate the extent to which each sensor-placement and reconstruction method reproduced its temporal evolution.

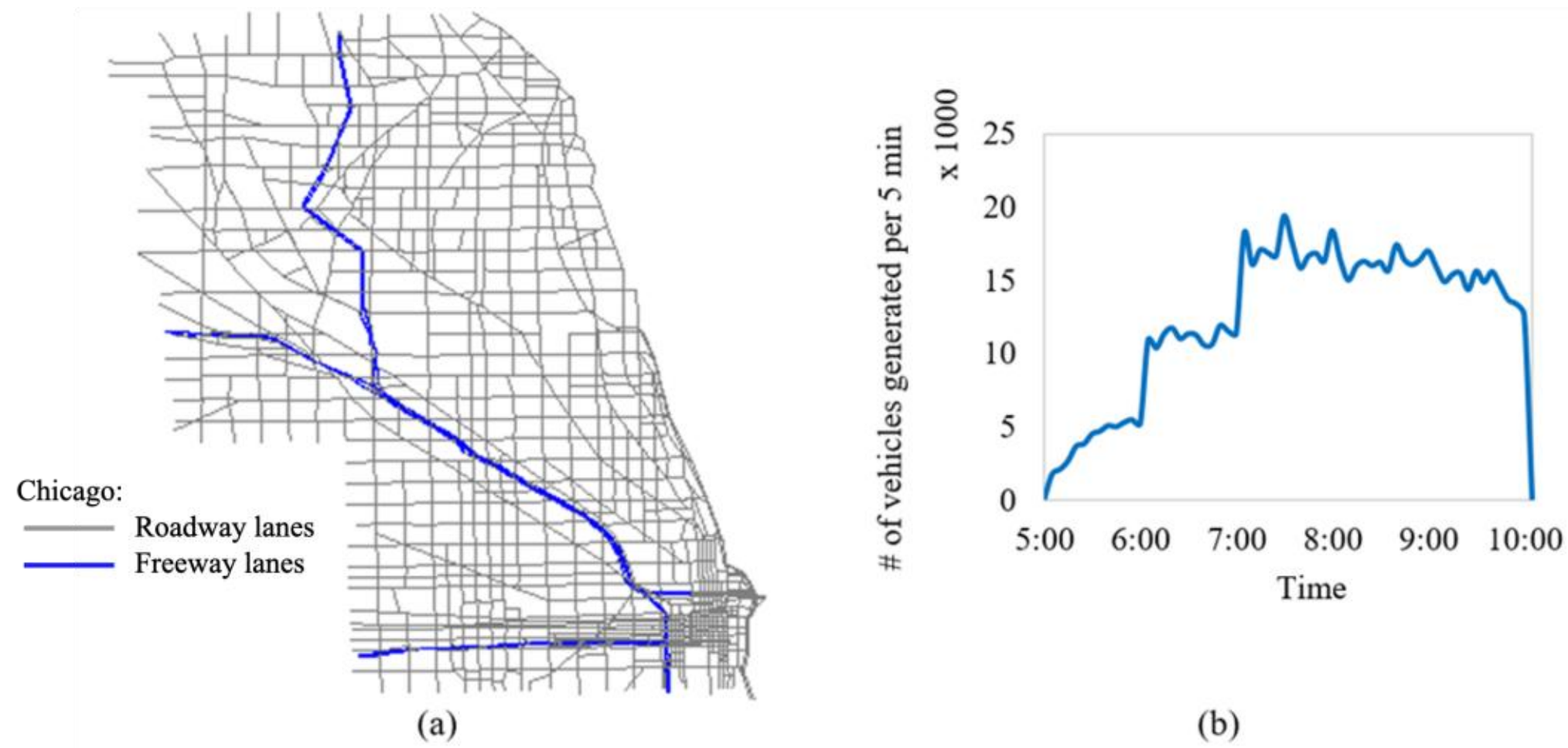

**Figure 1 Chicago roadway network and temporal demand profile used in the simulation (Fakhrmoosavi et al. 2023)**

## 4.2. Data Partitioning

The 420 time intervals were divided into training, validation, and testing subsets. To preserve short-term temporal dependencies and maintain a comparable distribution of network traffic conditions across the three subsets, every 12 consecutive one-minute intervals were grouped into a temporal block. This procedure produced 35 non-overlapping blocks, each representing 12 minutes of simulated traffic conditions. The blocks were assigned to the three subsets according to their network-average density distributions so that each subset included observations spanning the range of uncongested, transitional, and congested conditions represented in the simulation. The final partition consisted of 15 training blocks, 10 validation blocks, and 10 testing blocks, corresponding to 180, 120, and 120 time intervals, respectively. All observations within a block were assigned to the same subset to prevent adjacent intervals from being divided across training, validation, and testing data.

The training subset was used to estimate the link-specific density and flow quantile thresholds, calculate the normalization statistics, construct the NMI-based coverage and redundancy matrices, select sensor locations, and optimize the reconstruction model. The validation subset was used for early stopping, model selection, and estimation of the post-processing calibration coefficients. The testing subset was excluded from all sensor-selection, model-training, and calibration procedures and was used only for final performance evaluation.

### 4.3. Implementation Settings

Following the discretization procedure described in the methodology for NMI estimation, both density and flow were partitioned into four link-specific quantile bins, resulting in up to 16 joint traffic states for each link. In NMI computation, the maximum temporal lag was set to 1 minute. In sensor selection, the link-importance coefficient and redundancy penalty was set to 0.01 and 0.005, respectively. For reconstruction-model training, the observed-link loss coefficient, MFD loss coefficient, and physical penalty coefficient were set to 0.1, 0.2, and 0.2, respectively. The temporal input window contained four consecutive one-minute intervals, including the current interval and the three preceding intervals. The graph diffusion order was set to three, meaning that each graph diffusion operation aggregates information from neighboring roadway links connected through paths of up to three directed link-to-link movements. Information from more distant links can still influence the final reconstruction through successive graph diffusion layers. The model used two bidirectional graph diffusion layers with a hidden dimension of 64, followed by a node-level linear output layer that estimated density and flow for every link.

Model parameters were optimized using the Adam optimizer with a learning rate of 0.001 and a mini-batch size of 32. Training intervals were randomly shuffled before mini-batch formation in each epoch. Model selection was based on the total validation loss, and the parameter values associated with the minimum validation loss were retained. The same network representation, temporal inputs, loss-function coefficients, training procedure, and post-processing checks were applied across all sensor-placement scenarios, so the model performance could be evaluated under a consistent reconstruction-model configuration.

## 5. Results and Discussion

The performance of the proposed NMI-IGNNK framework was evaluated by examining its ability to select informative sensor locations and reconstruct both link-level traffic states and the resulting network MFD across a range of sensor-coverage levels. The analysis compared reconstructed and ground-truth network-average flow and density for the independent testing subset and for the complete simulation period. Performance was assessed using flow RMSE, density RMSE, and joint normalized flow–density RMSE, together with visual comparisons of the reconstructed and ground-truth MFDs. To determine whether the proposed information-theoretic and graph-based framework provides advantages over an established dimensionality-reduction approach, the results were also compared with a PCA-based sensor-selection and reconstruction baseline. The PCA approach identifies dominant traffic patterns, with sensor locations selected as the links having the largest absolute loadings in the retained principal components (Saffari et al. 2020). Multi-linear regression models are then used to estimate the retained principal components from measurements at the selected links, after which the full-network link-level density and flow states are reconstructed using the PCA projection matrices and aggregated to generate the MFD. The comparison examines reconstruction accuracy over a range of sensor ratios considered as a representative of sensor budget, the ability of each method to reproduce the network MFD, and how performance changes as additional sensors are introduced. Here, we report NMI-IGNNK results for sensor ratios ranging from 0.5% to 30%, whereas the PCA baseline is evaluated only over the smaller range, from 0.5% to 3%, supported by the rank of the traffic-state matrices.

Figure 2 compares the reconstructed MFD obtained using the proposed NMI-IGNNK framework with the ground-truth MFD for the independent testing subset across different sensor ratios. At the lowest sensor ratios (0.5% and 1%), the reconstructed MFD captures the general shape of the traffic-state relationship but exhibits noticeable deviations, particularly in the initial low-flow/low-density period and the congested regime. Increasing the sensor ratio to 2% and 3% substantially improves the performance, with the reconstructed traffic states more closely following both branches of the MFD. Further increases to 5% and 10% reduce the remaining discrepancies, particularly along the high-density congested regime, while preserving the hysteresis pattern. At higher sensor ratios (i.e., 20% and 30%), the reconstructed MFDs show no clear further improvement. These results indicate that the proposed information-theoretic sensor placement and graph-based reconstruction framework accurately captures the network-level traffic state when sufficient sensor coverage is available, but further increases in sensor coverage do not necessarily improve reconstruction accuracy under the baseline parameter configuration.

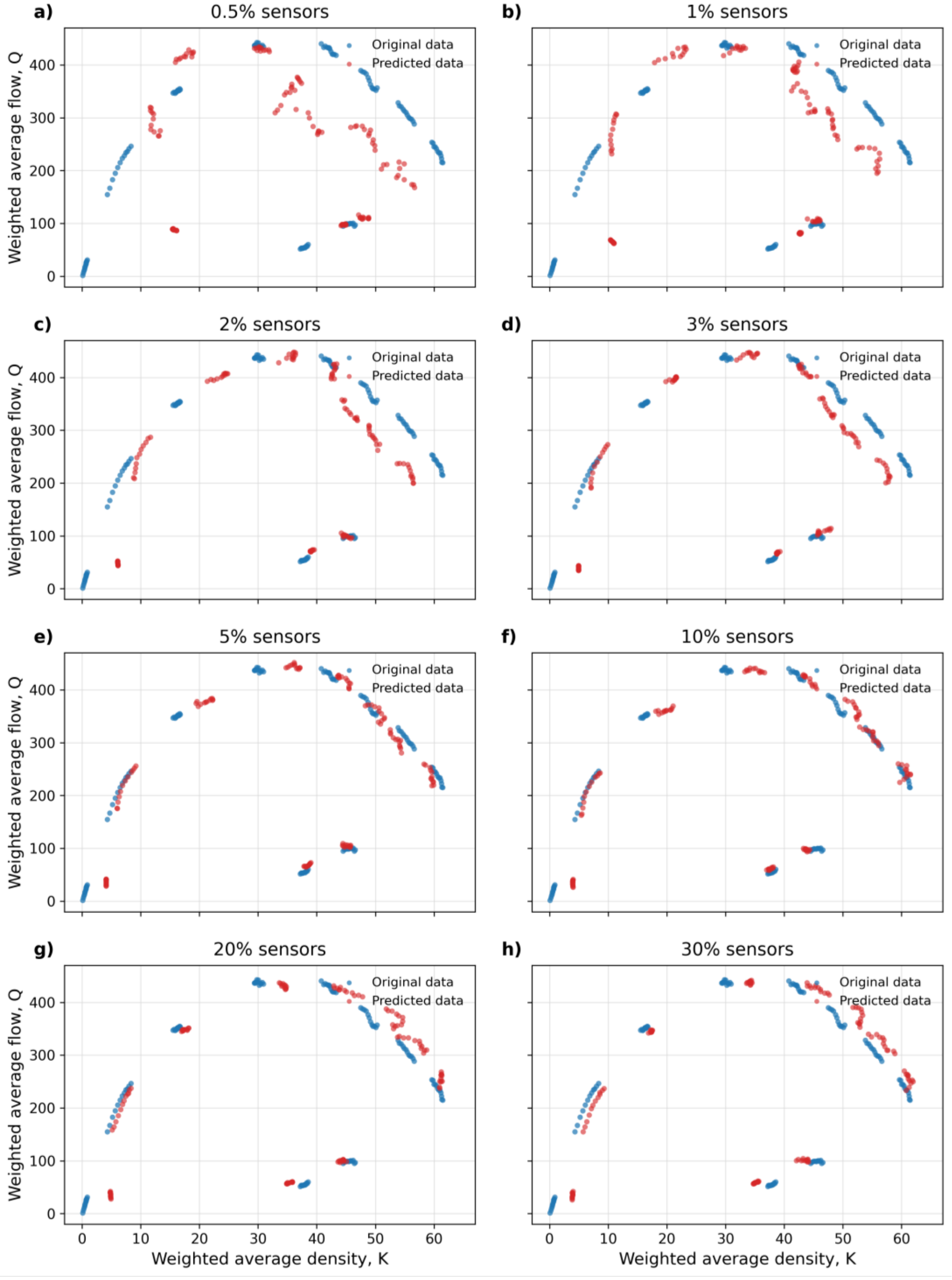

**Figure 2 Observed and predicted MFDs for the testing subset using the proposed NMI-IGNNK framework under different sensor ratios**

Figure 3 presents the reconstructed MFDs obtained using the PCA-based sensor-selection method for the testing subset. At sensor ratios between 0.5% and 2%, the PCA-based method reproduces the overall shape of the observed MFD but exhibits substantial scatter, particularly along the congested part and during the transition from uncongested to congested regimes. Increasing the sensor ratio from 0.5% to 2% provides considerable improvements in reconstruction quality, while the 3% case shows no visible reduction in scatter. These results suggest that the PCA-based method can recover the dominant low-dimensional traffic patterns but is less effective in preserving the detailed nonlinear structure of the network traffic states.

Unlike the NMI-IGNNK framework, the PCA-based approach cannot be applied to larger sensor ratios because the number of selected sensors is limited by the rank of the training traffic-state matrix. In this study, the density and flow matrices each contain 180 training time intervals and 4,805 directed links, giving a rank no greater than 180. The PCA method estimates one regression model for each retained principal component using measurements from the selected sensors. Consequently, the number of selected sensors cannot exceed the number of independent principal components. Sensor ratios above 3.75% correspond to more than 180 sensors ($0.0375 \times 4{,}805 \approx 180$). In contrast, the proposed NMI-IGNNK framework does not rely on low-rank matrix decomposition and remains applicable over substantially larger sensor ratios.

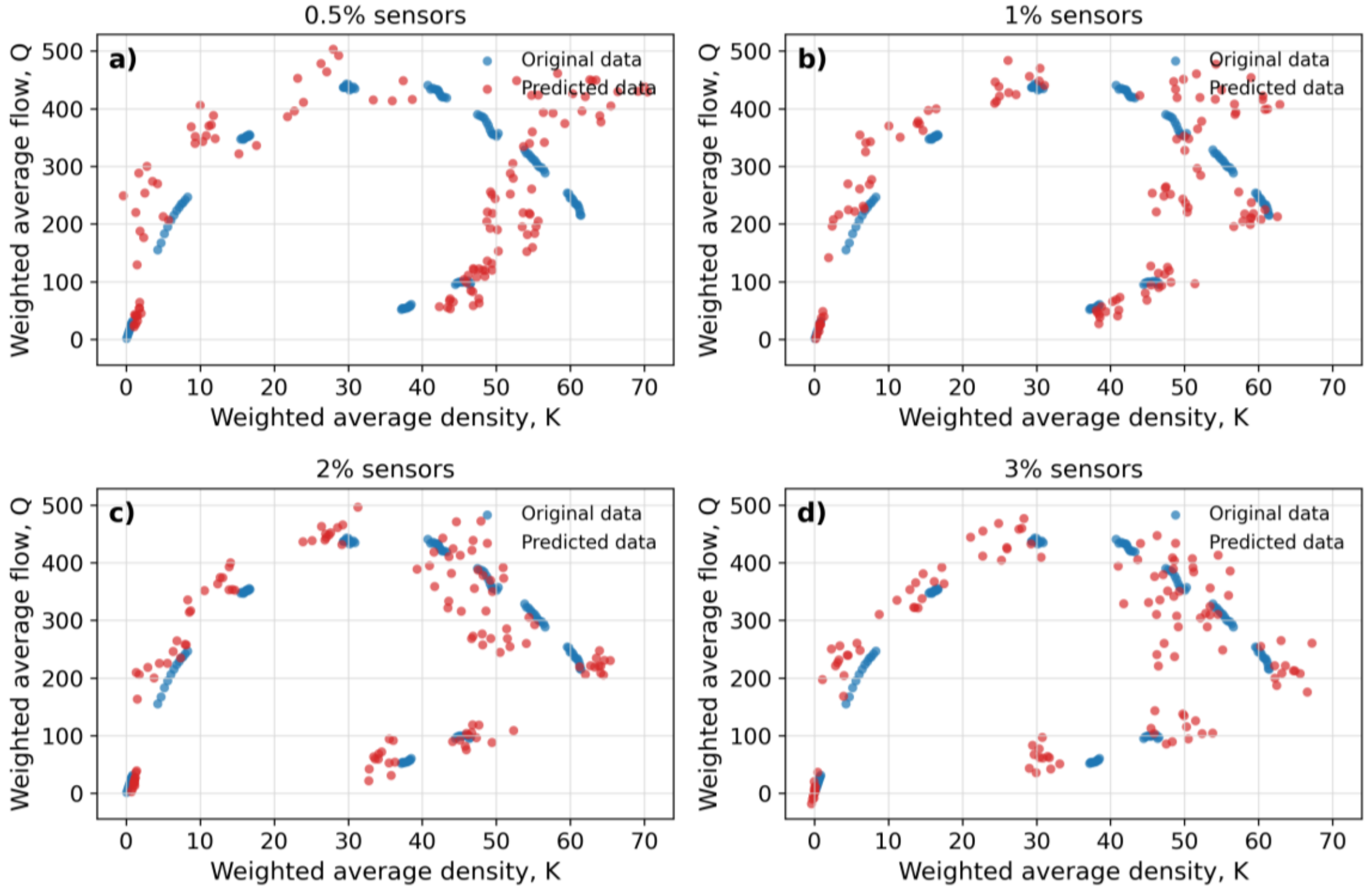

**Figure 3 Observed and predicted MFDs for the testing subset using the PCA-based sensor-selection and reconstruction method**

Figures 4 and 5 compare the observed MFD with the MFDs derived from traffic states estimated by the NMI-IGNNK and PCA-based methods, respectively, using all available time intervals. For NMI-IGNNK, the estimated MFD deviates substantially from the observed MFD at small sensor ratios (i.e., 0.5% and 1%), particularly along the congested part. The deviations decrease significantly at 2% and 3%, and the method reproduces the overall MFD more accurately as the sensor ratio increases. By comparison, the PCA-based estimates show substantial scatter at all evaluated sensor ratios, particularly near capacity and along the congested regime. Although PCA captures the general MFD shape, its performance does not consistently improve from 0.5% to 3%. Overall, the results indicate that NMI-IGNNK benefits more consistently from additional sensors and more accurately reproduces the network-level flow–density relationship.

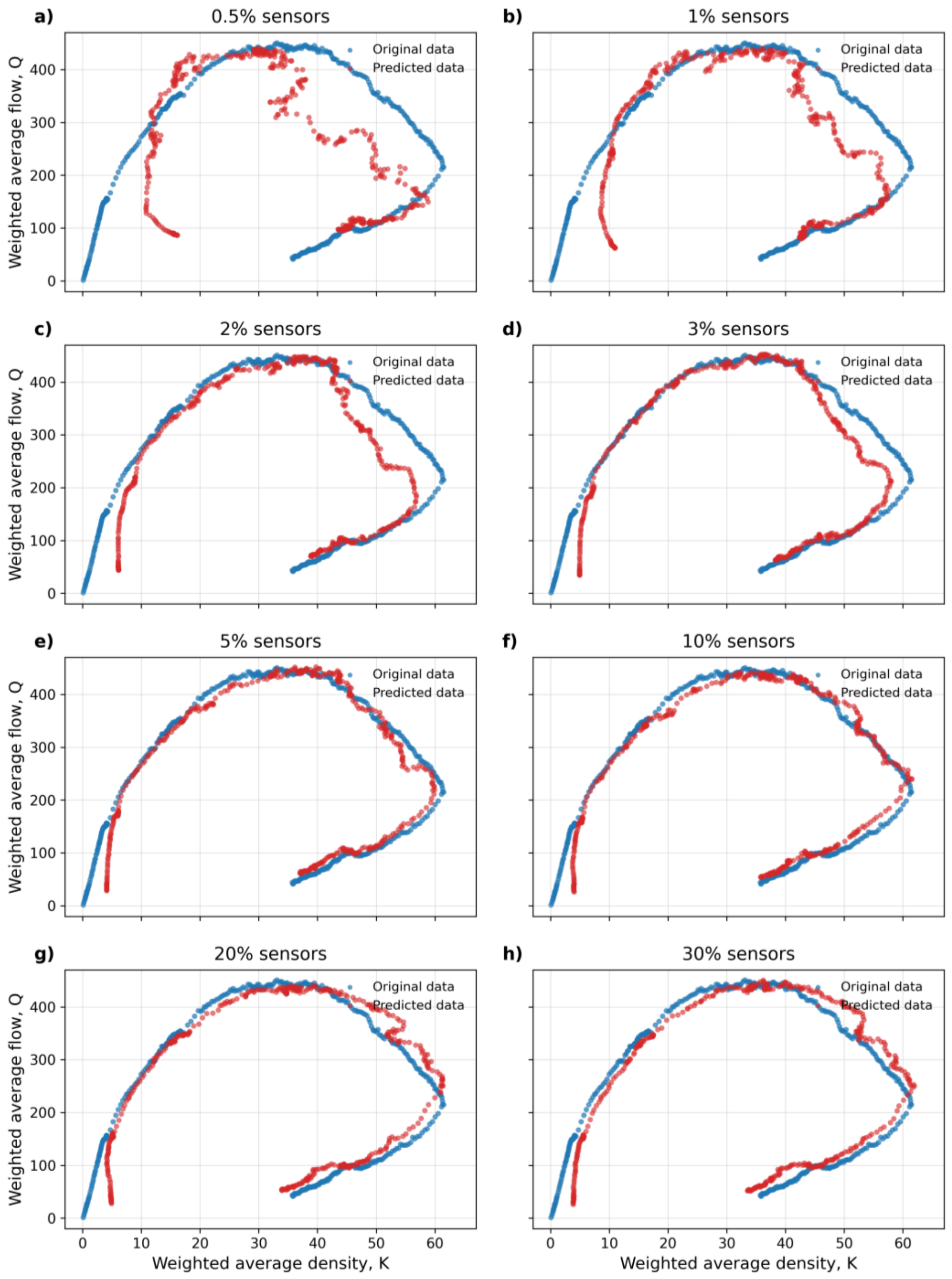

**Figure 4 Observed and NMI-IGNNK-estimated MFDs for the complete analysis period at different sensor ratios**

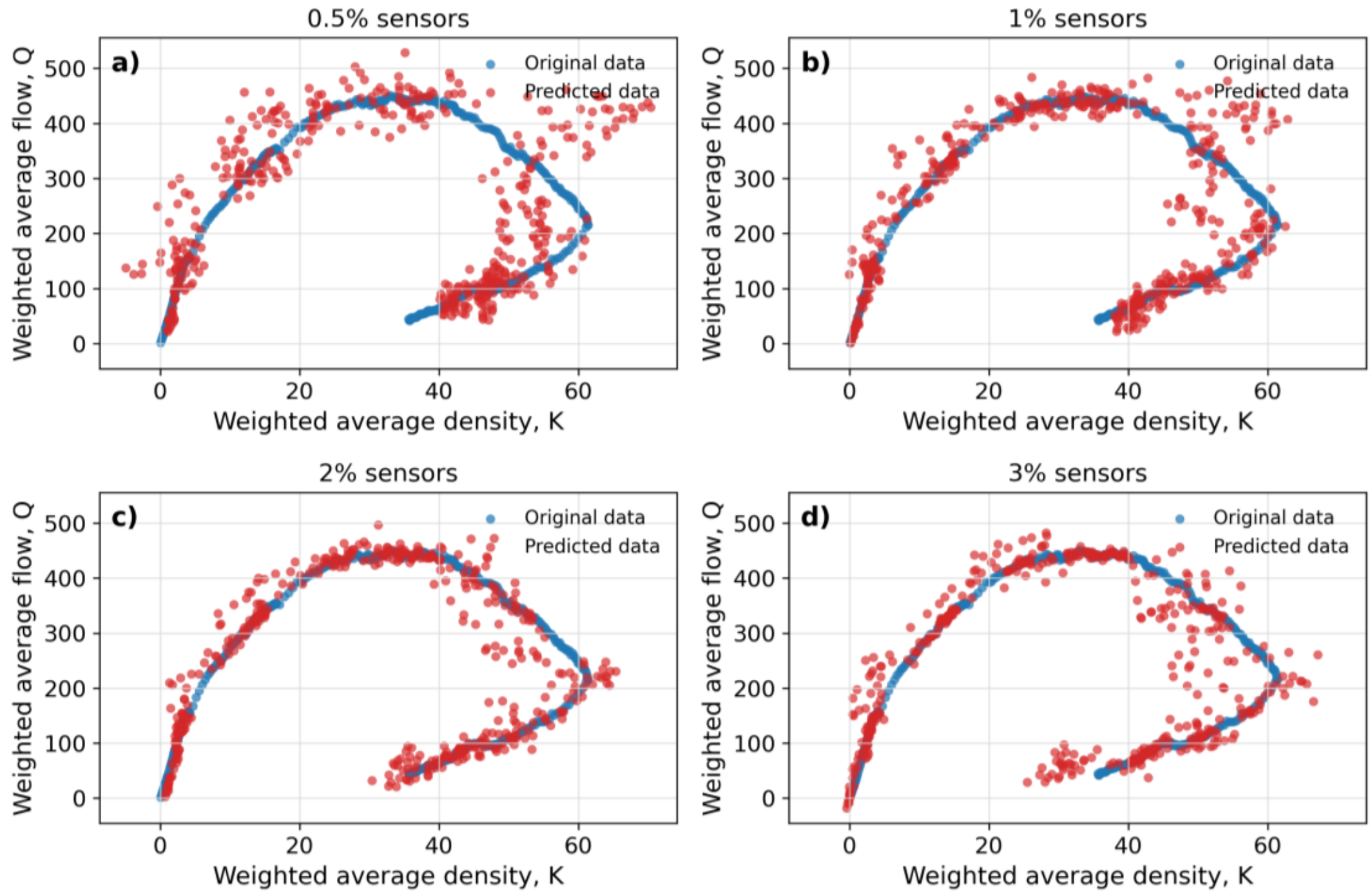

**Figure 5 Observed and PCA-estimated MFDs for the complete analysis period at different sensor ratios**

To quantify the differences in the observed and predicted MFDs using NMI-IGNNK and PCA-based sensor placement methods, Figure 6 reports the flow RMSE, density RMSE, and normalized joint RMSE for the testing subset and the complete analysis period, respectively. Over the directly comparable sensor ratios of 0.5%–3%, PCA generally produces lower RMSE values from 0.5% to 2%, while NMI-IGNNK achieves lower RMSE for all measurements at 3% for both the testing and all simulation intervals. For the testing subset, the normalized joint RMSE values are 0.072 for NMI-IGNNK and 0.100 for PCA for normalized joint RMSE. For all simulation intervals, the normalized joint RMSE is 0.064 for NMI-IGNNK and 0.075 for PCA. These differences correspond to reductions of 28.1% and 14.3% in normalized joint RMSE for the testing subset and all intervals, respectively. The PCA errors reach their minimum at a 2% sensor ratio and then increase at 3%. For the testing subset, its normalized joint RMSE decreases from 0.165 at 0.5% to 0.079 at 2% before increasing to 0.100 at 3%. The same pattern appears for all intervals. Thus, adding sensors from 2% to 3% does not improve the PCA results.

NMI-IGNNK errors exhibit an overall downward trend as the sensor ratio increases, despite minor fluctuations at higher sensor ratios. From 0.5% to 3%, all three RMSE measures decrease considerably, with normalized joint RMSE falling from 0.178 to 0.072 for the testing subset and from 0.156 to 0.064 for all simulation intervals. As the sensor ratio increases from 3% to 10%, normalized joint RMSE continues to decline, reaching 0.046 and 0.043 for the testing subset and all intervals, respectively. These values represent reductions of 35.7% and 33.0% relative to the 3% results. At 20% and 30%, testing-subset normalized joint RMSE no longer decreases and instead rises slightly to 0.052 and 0.050, respectively, while the corresponding all-interval values remain near a plateau at 0.044. Overall, PCA generally achieves lower flow and normalized joint RMSE at sensor ratios of 0.5%–2%, whereas NMI-IGNNK achieves better performance across all three metrics at 3% and benefits substantially from further increases in sensor coverage up to 10%. As discussed previously, the largest deviations for NMI-IGNNK at low sensor ratios occur during the initial low-flow, low-density period and under congested conditions. Nevertheless, for most sensor ratios, NMI-IGNNK produces less dispersion around the observed flow–density relationship than the PCA-based method.

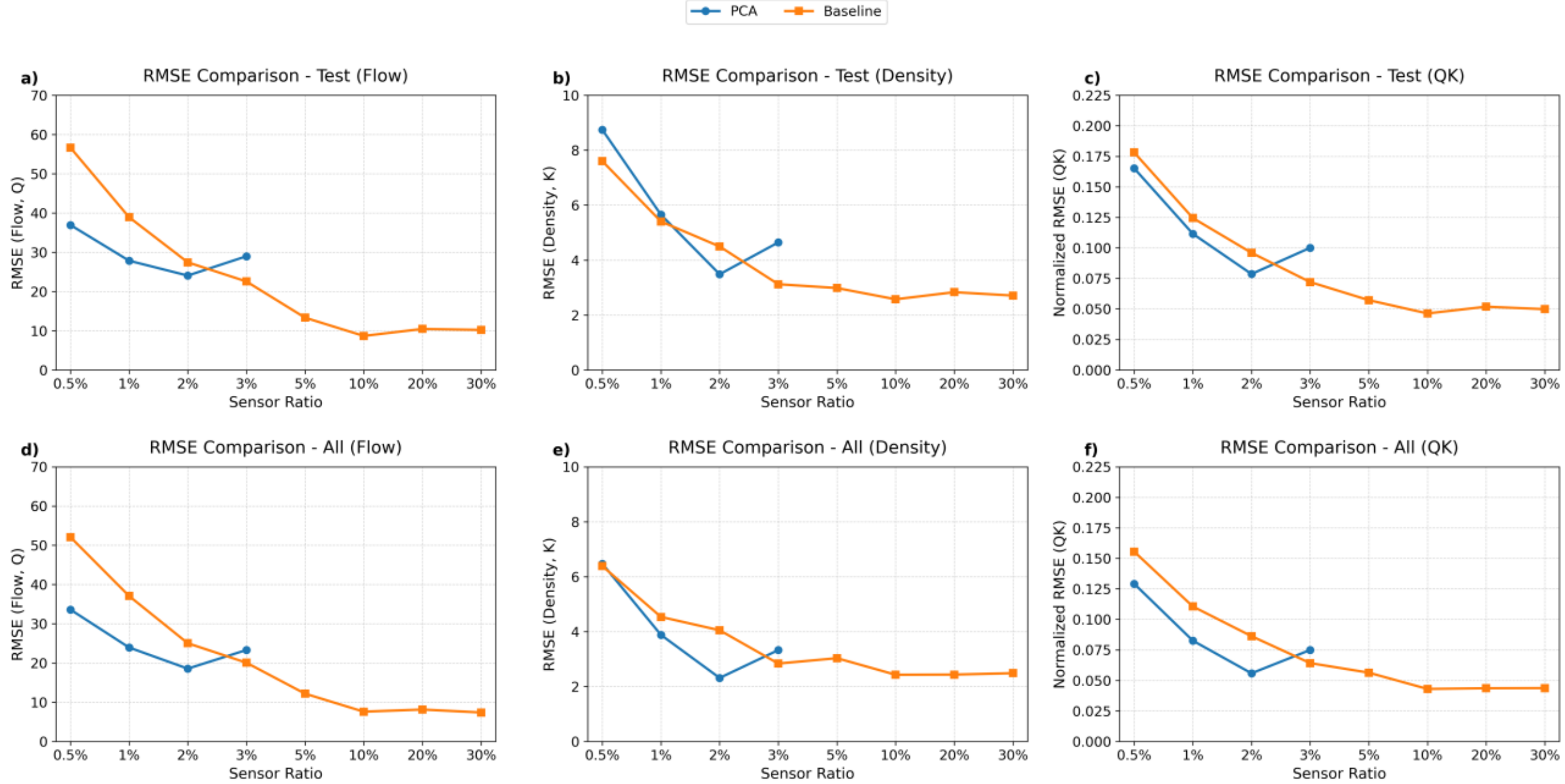

**Figure 6 Comparison of flow, density, and normalized joint RMSE between the proposed NMI-IGNNK framework and the PCA-based method**

The comparison between the NMI-IGNNK and PCA methods indicates that their relative performance varies substantially with the available sensor budget. Under low-budget conditions, defined here as sensor coverage below 3% of the network links, the PCA baseline generally produced lower errors in flow, density, and joint RMSE. This result suggests that PCA can effectively exploit the dominant low-rank structure of the traffic-state matrices and reconstruct the principal patterns of the original network conditions using only a limited number of sensors. However, the PCA results showed some nonmonotonic variation as higher sensor coverage was considered. Furthermore, the maximum number of independent sensor locations supported by the PCA formulation was constrained by the rank of the traffic-state matrix and, therefore, by the number of available time intervals.

In contrast, the NMI-IGNNK framework remained applicable over a wide range of sensor budgets, and its reconstruction errors decreased as the number of selected sensors increased. This pattern indicates that the information-theoretic sensor-selection strategy identifies links that provide complementary information about unobserved locations, while the graph-based reconstruction model uses the roadway topology to propagate this information across the network. Together, these characteristics enable the framework to scale to substantially larger sensor deployments and progressively improve network-level traffic-state reconstruction. Therefore, while PCA provides an effective baseline for highly constrained sensor budgets, NMI-IGNNK is more suitable for applications requiring higher reconstruction accuracy and greater flexibility in sensor deployment.

## 6. Sensitivity Analysis

To evaluate the robustness of the proposed information-theoretic sensor-placement procedure, we conducted sensitivity analyses for four parameters that directly influence sensor selection: the number of quantile bins used to construct joint flow–density states, the maximum temporal lag considered in NMI estimation, as well as the link-importance coefficient and redundancy-penalty coefficients in the candidate score calculation. These parameters were selected because they affect the information representation, temporal dependence, and tradeoff for candidate sensors, respectively, and can therefore alter the selected sensor locations. Other parameters associated with the IGNNK reconstruction model, such as the temporal input window, diffusion order, hidden dimension, and optimization settings, were not emphasized because they affect the reconstruction stage after sensor locations are determined and do not directly define the proposed information-theoretic placement criterion. The different settings for all tested hyperparameters are shown in Table 1.

**Table 1. Sensitivity analysis settings to information-theoretic sensor-selection parameters**

| Parameter | Baseline | Setting 1 | Setting 2 | Setting 3 |
|---|---|---|---|---|
| Number of quantile bins | 4 | 2 | 3 | 5 |
| Maximum temporal lag, $\tau_{\max}$ (min) | 1 | 0 | 3 | 5 |
| Link-importance coefficient, $\gamma$ | 0.01 | 0.005 | 0.02 | 0.04 |
| Redundancy penalty, $\alpha$ | 0.005 | 0.0025 | 0.015 | 0.025 |

To investigate the effects of different settings for these hyperparameters under different budget scenarios, we conducted a sensitivity analysis under both the 3% and 10% sensor ratios, representing low- and high-budget scenarios, respectively. To evaluate the performance of each setting, we used the normalized joint RMSE to evaluate the model accuracy, and the Jaccard index to measure the similarity between a tested sensor set, $\mathcal{S}$, and the same-ratio baseline set, $\mathcal{S}_b$:

$$J(\mathcal{S}, \mathcal{S}_b) = \frac{|\mathcal{S} \cap \mathcal{S}_b|}{|\mathcal{S} \cup \mathcal{S}_b|} \tag{27}$$

where $J(\mathcal{S}, \mathcal{S}_b)$ ranges from zero for disjoint sets to one for identical sets.

The performance of these settings under 3% and 10% sensor ratio is shown in Figures 7 to 10, while the similarity of the sensor sets to the baseline setting is shown in Table 2. As shown in the figures, alternative settings for the number of quantile bins, maximum temporal lag, and link-importance coefficient improved performance relative to the baseline in some cases. However, among the configurations evaluated, no single fixed combination of the four hyperparameters achieved the lowest normalized joint RMSE for both the testing subset and all intervals at both sensor ratios. This indicates that the hyperparameters perform differently under different budget levels.

**Table 2. Sensitivity of selected sensor locations to information-theoretic sensor-selection parameters under 3% and 10% sensor**

| Parameter | Sensor Ratio | Baseline | Setting 1 | Setting 2 | Setting 3 |
|---|---|---|---|---|---|
| Number of quantile bins | | 4 | 2 | 3 | 5 |
| Sensor-set similarity (Jaccard) | 3% | 1.000 | 0.518 | 0.534 | 0.611 |
| | 10% | 1.000 | 0.402 | 0.498 | 0.552 |
| Maximum temporal lag, $\tau_{\max}$ (min) | | 1 | 0 | 3 | 5 |
| Sensor-set similarity (Jaccard) | 3% | 1.000 | 0.835 | 0.835 | 0.801 |
| | 10% | 1.000 | 0.673 | 0.825 | 0.775 |
| Link-importance coefficient, $\gamma$ | | 0.01 | 0.005 | 0.02 | 0.04 |
| Sensor-set similarity (Jaccard) | 3% | 1.000 | 0.667 | 0.758 | 0.667 |
| | 10% | 1.000 | 0.740 | 0.691 | 0.475 |
| Redundancy penalty, $\alpha$ | | 0.005 | 0.0025 | 0.015 | 0.025 |
| Sensor-set similarity (Jaccard) | 3% | 1.000 | 0.986 | 0.847 | 0.706 |
| | 10% | 1.000 | 0.727 | 0.508 | 0.413 |

Figure 7 further shows that under both the 3% and 10% sensor ratios, a higher number of bins generally resulted in better performance (Bins = 4 or 5). Specifically, the $RMSE\ (Q, K)$ decreased by about 13.3% and 14.2% from 2 to 4 quantile bins for the test and all subsets, respectively, under the 3% sensor ratio, while the corresponding decreases were 20.6% and 22.2% under the 10% sensor ratio. Under both ratios, as the number of quantile bins increases, the $RMSE\ (Q, K)$ consistently decreases until the number of bins reaches 4, while the performance difference between 4 and 5 bins is not considerable. This indicates that a relatively higher number of quantile bins may help improve the precision of measurement and the performance of the NMI-IGNNK model. However, once the number of quantile bins reaches a certain level, further improvement is subtle. It is also worth noting that, as the Jaccard values for each setting show, changes in the number of quantile bins, including both increases and decreases, considerably change the similarity between sensor sets, which means that the number of quantile bins can substantially affect sensor selection.

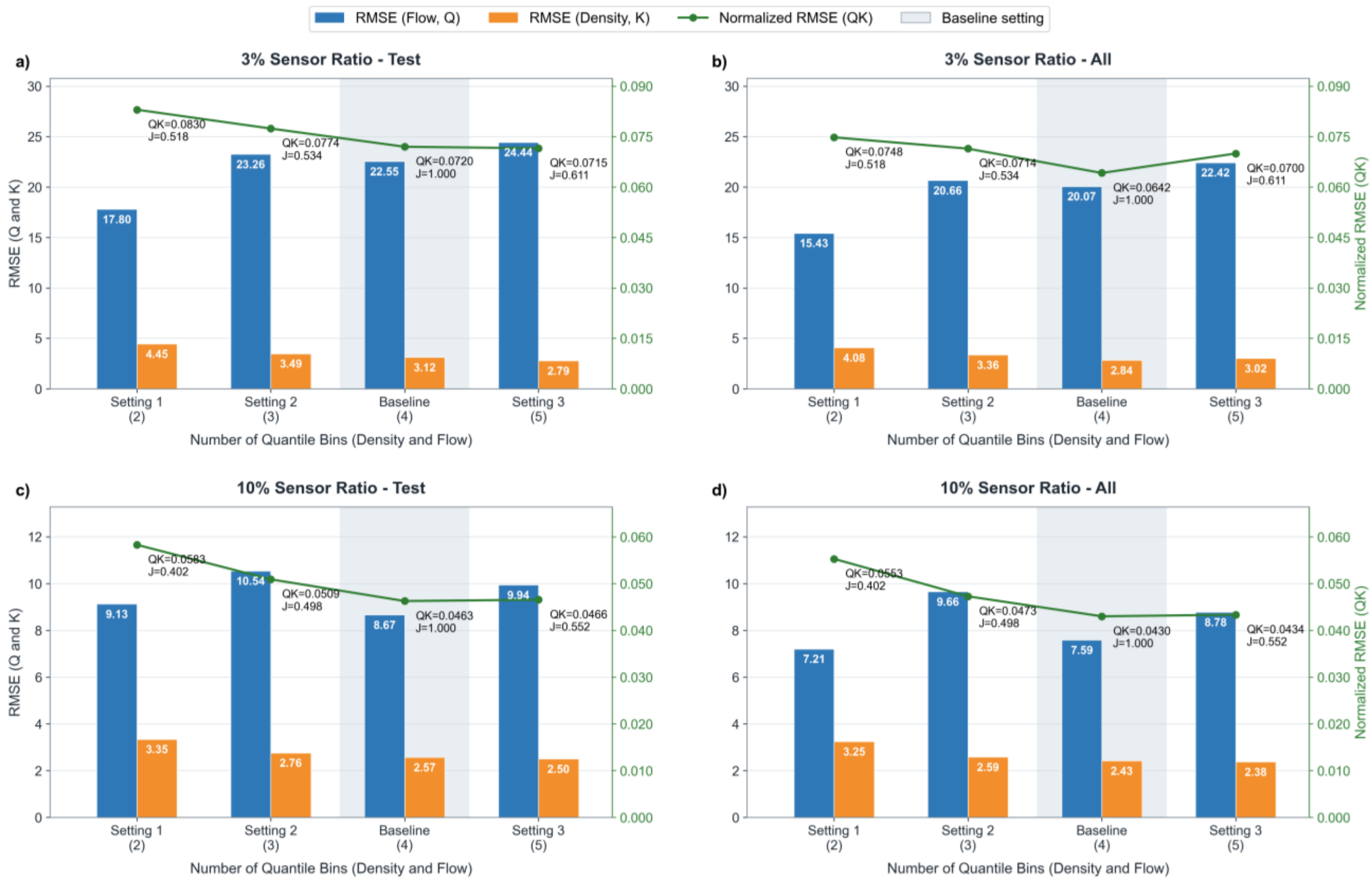

**Figure 7 Sensitivity to the number of quantile bins for density and flow**

Figure 8 shows the sensitivity of the results to the maximum temporal lag parameter. As can be seen, under the lower sensor ratio (3%), a lower maximum temporal lag resulted in better RMSE value, while under the higher sensor ratio (10%), the performance fluctuated across different maximum temporal lags. Specifically, the $RMSE\ (Q, K)$ increased by about 17.0% and 14.0% from a maximum temporal lag of 0 to 5 for the test and all subsets, respectively, under the 3% sensor ratio. As the maximum temporal lag essentially determines the temporal lag considered by the NMI matrix, this result may indicate that under lower sensor ratio scenarios, a higher lag-aware scheme may be less important, while under higher sensor ratio scenarios, once the information coverage is relatively high, the performance differences across different temporal lags are not consistent. Under the lower sensor ratio, incorporating longer temporal dependencies may shift the sensor selection toward locations with stronger lagged relationships but potentially weaker direct representation of the current network state. When only a limited number of sensors are available, such a trade-off may reduce reconstruction performance, making shorter temporal lags more effective. In contrast, under the higher sensor ratio, the increased spatial coverage may compensate for differences in sensor selection caused by different temporal lags, resulting in a weaker and non-monotonic effect on reconstruction performance. However, the Jaccard values indicate that the maximum temporal lag considerably influences the composition of the selected sensor set. Among the tested settings, larger deviations from the baseline in the same direction are associated with lower similarity to the baseline sensor set.

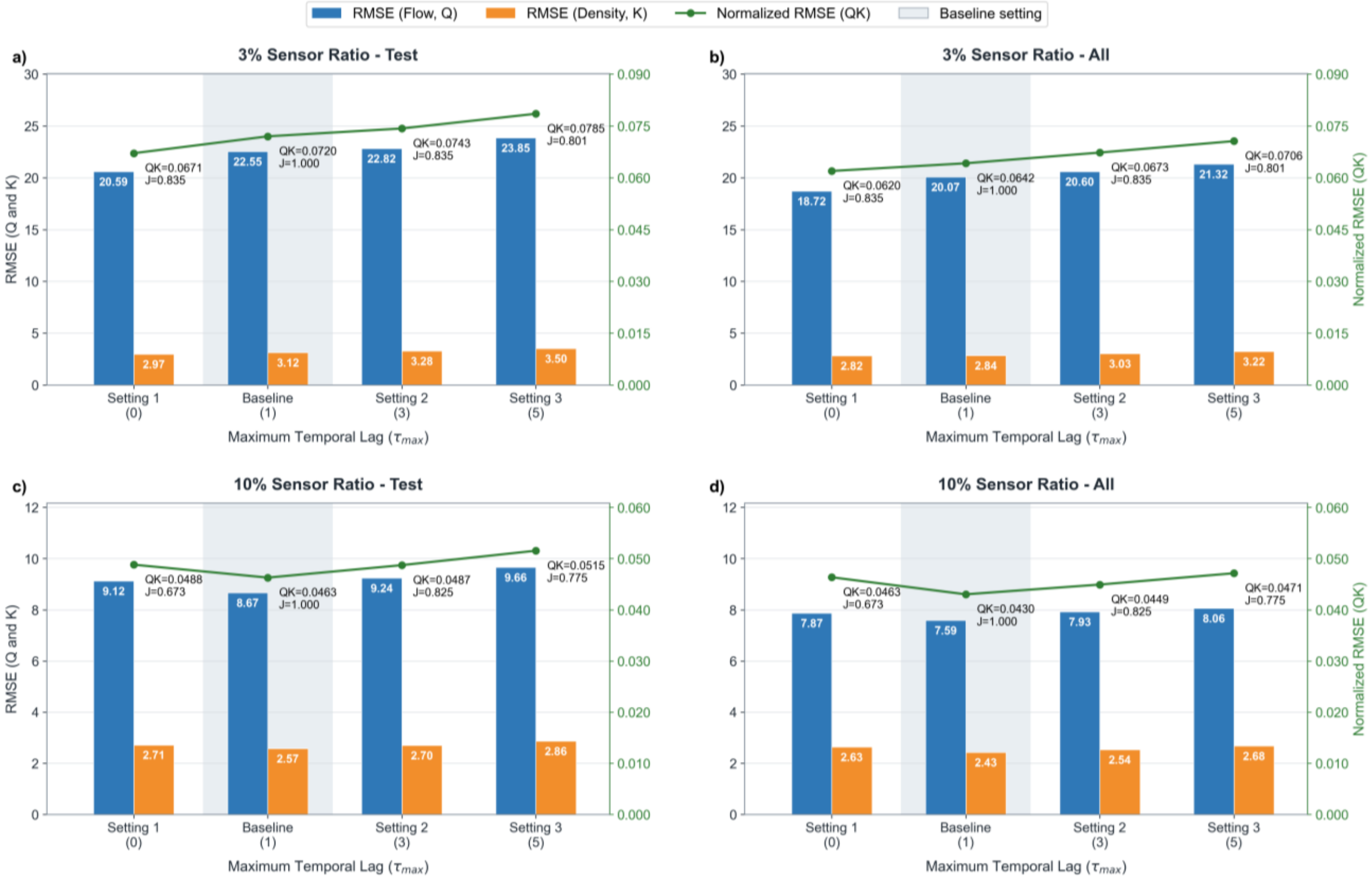

**Figure 8 Sensitivity to maximum temporal lag, $\tau_{\max}$**

Beyond the temporal dependence captured through the maximum lag, the link-importance coefficient controls the contribution of individual link characteristics to sensor selection. As Figure 9 shows, under the lower sensor ratio (3%), a lower link-importance coefficient resulted in better performance. In contrast, under the higher sensor ratio (10%), a higher link-importance coefficient resulted in better performance for the test subset. Specifically, the $RMSE\ (Q, K)$ increased considerably by about 37.2% and 44.4% as the link-importance coefficient increased from 0.005 to 0.04 for the test subset and all intervals, respectively, under the 3% sensor ratio, while it decreased by about 14.3% for the test subset under the 10% sensor ratio. The link-importance coefficient in the candidate score equation determines the importance assigned to the link itself, based on the lane length, mean traffic activity, and temporal variability on that link. Under a lower sensor ratio, coverage gain is the primary factor in sensor selection and reconstruction, and therefore, a relatively low link-importance coefficient is preferred. Under a higher sensor ratio, once the information coverage reaches a higher level, the link-importance of the selected links become more important; therefore, a higher link-importance coefficient is suggested.

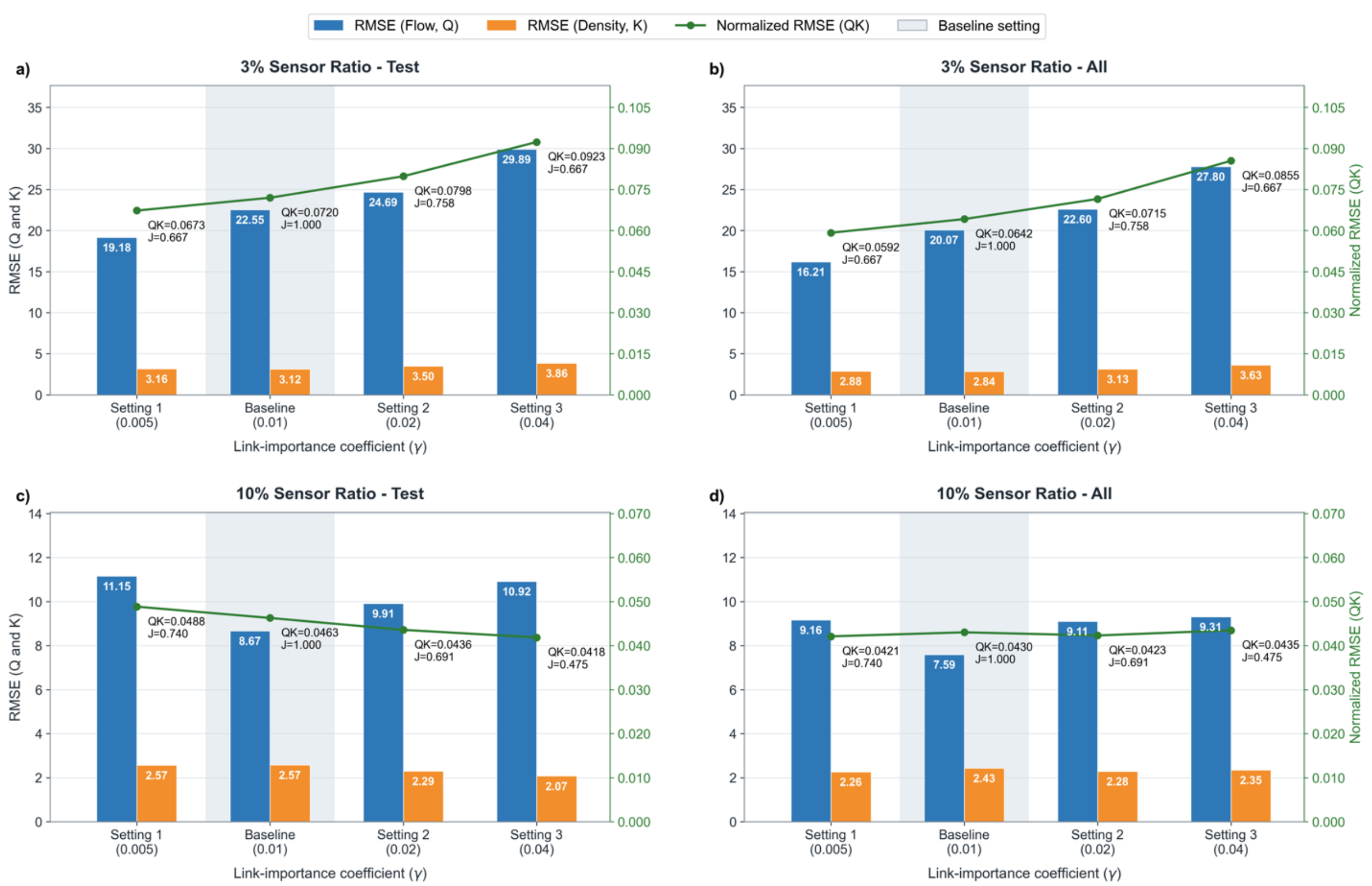

**Figure 9 Sensitivity to link-importance coefficient, $\gamma$**

To evaluate how the redundancy penalty affects reconstruction performance under different sensor budgets, Figure 10 compares the tested $\alpha$ values at sensor ratios of 3% and 10%. At 3% sensor coverage, reconstruction performance varied only modestly across the tested values, with no consistent trend. However, at 10% coverage, the normalized joint RMSE increased by approximately 73.0% and 57.6% for the testing subset and all intervals, respectively, as $\alpha$ increased from 0.005 to 0.025. The redundancy penalty in the candidate score equation reduces redundancy in the selected sensor set. Similar to the link-importance coefficient, under a lower sensor ratio, coverage gain is the primary factor in sensor selection and reconstruction, while the redundancy penalty does not have a considerable impact. Under a higher sensor ratio, as the information coverage reaches a higher level, redundancy becomes more considerable. Therefore, the selection in redundancy penalty shows considerable differences. This may be because, as the selected sensor set grows, marginal coverage gains decrease and redundancy has a greater relative influence on the candidate score, making an excessive redundancy penalty more likely to over-penalize potentially informative sensor overlap. The sensor selection trace output also shows that under the 3% sensor ratio, the first 22 selected sensors remain the same across different settings of the redundancy penalty, while the sensors selected later begin to change. This may indicate that the importance of the redundancy penalty setting increases under higher sensor ratio scenarios.

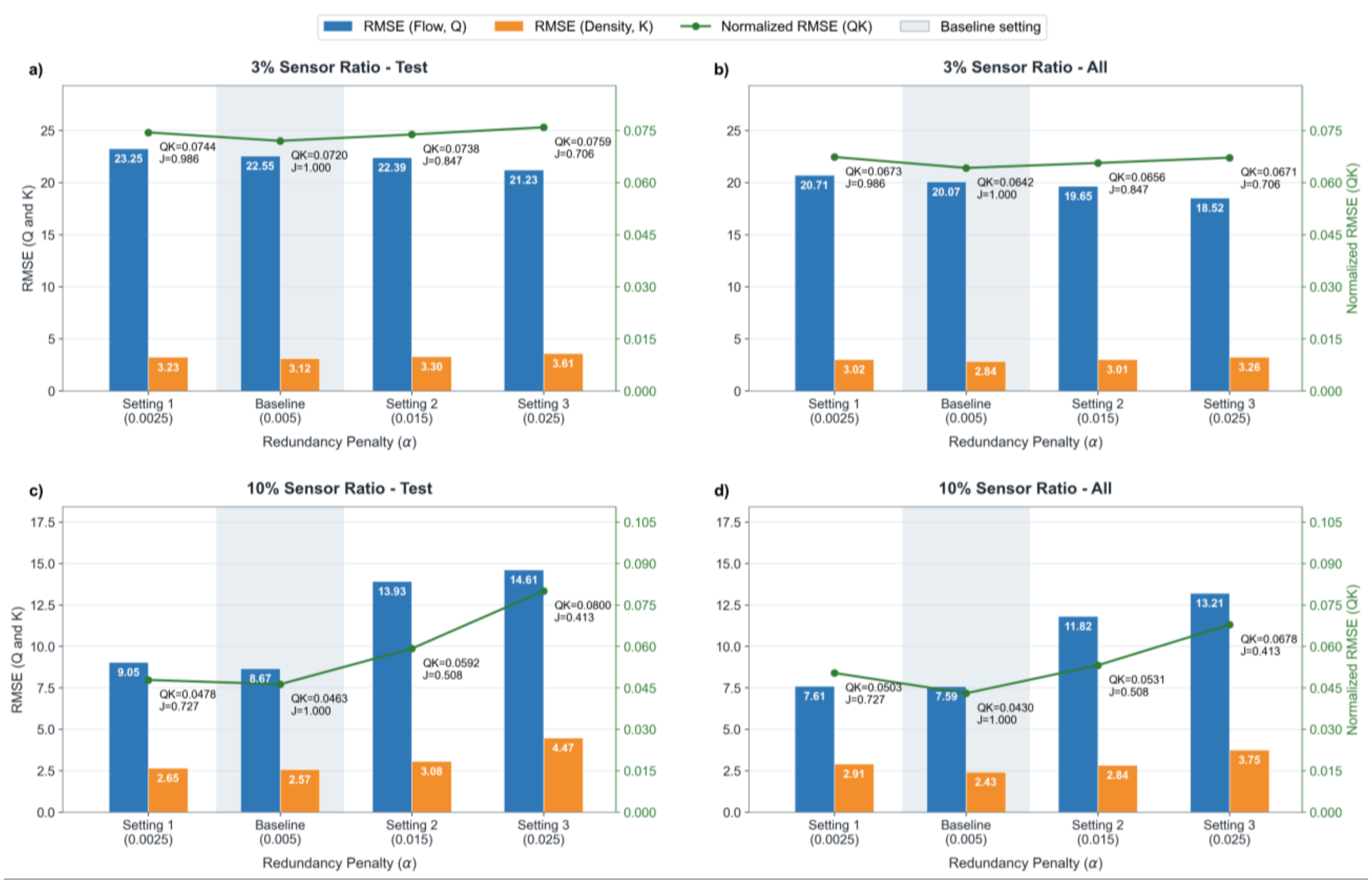

**Figure 10 Sensitivity to redundancy penalty, $\alpha$**

Overall, the sensitivity analysis shows that the effects of the four hyperparameters vary across different sensor budget levels. For the number of quantile bins, a relatively higher value generally improves NMI-IGNNK model performance under both sensor ratios, while the improvement becomes limited once the number of bins reaches approximately 4. For the maximum temporal lag, a lower value perform better under the lower sensor ratio, whereas no consistent trend is observed under the higher sensor ratio, suggesting that the benefit of incorporating longer temporal dependencies becomes less considerable as sensor coverage increases. The link-importance coefficient shows an opposite pattern across the two budget levels: a lower value is preferred under the lower sensor ratio, where coverage gain dominates sensor selection, while a higher value performs better under the higher sensor ratio, where the importance of the selected links themselves becomes more relevant after sufficient coverage is achieved. For the redundancy penalty, performance fluctuates under the lower sensor ratio, while a lower penalty is clearly preferred under the higher sensor ratio. These results indicate that while some hyperparameters, such as the number of quantile bins, show relatively consistent effects across budget levels, others are relatively budget-dependent and should be calibrated according to the sensor ratio.

## 7. Conclusions

This study developed NMI–IGNNK, an integrated framework for link-based traffic sensor placement and network traffic-state reconstruction. The sensor-selection component represents each link using joint flow–density states, quantifies directed lag-aware dependence between candidate and target links, and applies a greedy objective that maximizes weighted information coverage and link-importance while limiting redundancy among selected sensors. The selected links define a fixed observation mask for an IGNNK-based model that reconstructs flow and density at unobserved links through bidirectional graph diffusion while preserving measurements at selected links and generating the corresponding network MFD. The framework was evaluated using simulated traffic data from a large-scale Chicago network and compared with a PCA-based baseline over varying sensor ratios. NMI–IGNNK was evaluated at sensor ratios from 0.5% to 30% and compared with PCA evaluated from 0.5% to 3%, with the upper limit imposed by the rank of the training matrices.

When evaluated under varying sensor ratio, the results indicate that the performance of both methods depends on sensor coverage. Over the common evaluation range of 0.5%–3%, PCA generally produced lower RMSE values at 0.5%–2%, whereas NMI-IGNNK achieved lower RMSE across all three measures at 3% for both the testing subset and all interval simulations. At 3%, NMI-IGNNK reduced normalized joint RMSE relative to PCA by 28.1% and 14.3% for the testing subset and all intervals, respectively. However, PCA errors exhibited a nonmonotonic pattern. For the testing subset, the normalized joint RMSE decreased from 0.165 at a 0.5% sensor ratio to 0.079 at 2% before increasing to 0.100 at 3%; the same trend was observed for the complete analysis period. The visual MFD comparisons also showed persistent scatter near capacity and along the congested regime, suggesting that PCA captures the dominant low-dimensional traffic patterns but does not consistently preserve the detailed network flow–density relationship. In contrast, NMI–IGNNK remained applicable over the full sensor-ratio range of 0.5%–30% and showed a substantial improvement in reconstruction accuracy as sensor coverage increased. For the testing subset, the normalized joint RMSE decreased from 0.178 at 0.5% to 0.046 at 10%, while the corresponding all-interval value decreased from 0.156 to 0.043. The MFD comparisons showed progressively closer agreement with the observed flow–density relationship, including the hysteresis pattern, as sensor coverage increased. Unlike PCA, NMI–IGNNK is not constrained by the one-to-one correspondence between retained principal components and selected sensor locations and therefore remains applicable beyond the rank limit of the training matrices.

Sensitivity analyses at sensor ratios of 3% and 10% revealed budget-dependent effects of the sensor-selection hyperparameters. A relatively higher number of quantile bins generally improved performance at both sensor ratios. Shorter maximum temporal lags performed better at 3%, whereas the response at 10% was nonmonotonic, with a maximum lag of one minute yielding the lowest normalized joint RMSE. Lower link-importance coefficients favored reconstruction accuracy at 3%. The redundancy penalty had a stronger effect at 10% than at 3%, with excessive penalties substantially increasing normalized joint RMSE. The Jaccard comparisons further showed that changes in these hyperparameters altered the selected sensor sets, while the associated changes in reconstruction accuracy varied across settings.

These findings indicate that both the sensor-placement approach and its hyperparameter configuration should be guided by the available sensor budget and the reconstruction objective. By integrating lag-aware joint flow–density dependence, redundancy control, and directed graph propagation, the proposed framework identifies and selects links whose measurements provide complementary information for reconstructing traffic states both locally and across the network. The proposed NMI-IGNNK framework has a few limitations. First, the proposed framework was developed and evaluated using simulated traffic data, which provided complete link-level density and flow observations throughout the network. Complete network observations were required to train, validate, and objectively evaluate the framework against known ground truth. In practical applications, the framework is intended to operate using measurements collected only from the selected sensor locations. Future research should validate the framework using real-world traffic sensor data to assess its performance under operational conditions. Second, the current evaluation did not explicitly examine the effects of measurement noise, detector failures, or missing observations on sensor selection and reconstruction performance. Future work should investigate the performance under these conditions and develop strategies to improve its resilience to imperfect traffic measurements.

**Acknowledgements**

The authors used OpenAI’s ChatGPT for grammar and spelling review, language refinement, and codes debugging. All AI-assisted outputs were reviewed and verified by the authors, who retain full responsibility for the manuscript’s content, analysis, and conclusions.

**Declaration of Conflicting Interests**

The authors declared no potential conflicts of interest with respect to the research, authorship, and/or publication of this article.